\documentclass[prd,amsmath,showpacs,preprintnumbers]{revtex4}
\usepackage{epsfig}
\usepackage{graphicx}
\begin{document}

\newcommand{\ebox}[2]{\epsfxsize=#1 \epsfbox[10 30 560 590]{#2}}
\newcommand{\be}{\begin{equation}}
\newcommand{\ee}{\end{equation}}
\newcommand{\rf}[4]{{\em {#1}} {\bf #2}, #3 (#4)}
\newcommand{\PR}[3]{\rf{Phys.\ Rev.}{#1}{#2}{#3}}
\newcommand{\pr}{Phys.\ Rev.\ }
\newcommand{\slh}{\!\!\!\slash}

\newcommand{\physl}{Phys.\ Lett.}
\newcommand{\cm}{Commun.\ Math.\ Phys.}
\newcommand{\psibar}{\overline{\psi}}
\newcommand{\Lqcd}{\Lambda_{\text{QCD}}}
\newcommand{\oa}[1]{\ensuremath{{\cal O}(a^{#1})}}
\newcommand{\eref}[1]{(Eq.~\ref{#1})}
\newcommand{\tr}{\text{Tr}}
\newcommand{\cond}{\ensuremath{-\langle\psibar\psi\rangle}}
%


\preprint{\vbox{\hfill \rm ADP-03-106/T544}}

\title{Scaling behavior of the overlap quark propagator in 
Landau gauge}

\author{Jianbo Zhang}
\author{Patrick O.\ Bowman${}^\dagger$} 
\author{Derek B.\ Leinweber}
\author{Anthony G.\ Williams}
\affiliation{CSSM Lattice Collaboration, \\
Special Research Center for the Subatomic Structure of
Matter (CSSM) and Department of Physics,
University of Adelaide 5005, Australia \\
${}^\dagger$Nuclear Theory Center, Indiana University, 
Bloomington Indiana 47405}

\author{Fr\'{e}d\'{e}ric D.R.\ Bonnet}
\affiliation{University of Regina, Department of Physics, 
University of Regina, Regina, SK, S4S 0A2, Canada}

\date{\today}

\begin{abstract}
The properties of the momentum space quark propagator in Landau gauge
are examined for the overlap quark action in quenched lattice QCD.
Numerical calculations are done on three lattices with different lattice
spacings and similar physical volumes to explore the approach of the quark
propagator toward the continuum limit.  We have calculated the
nonperturbative momentum-dependent wave function renormalization function $Z(\zeta^2;p)$
and the nonperturbative mass function $M(p)$ for a variety of bare quark 
masses and perform an extrapolation to the chiral limit.
We find the behavior of $Z(\zeta^2;p)$ and $M(p)$ are in reasonable agreement
between the two finer lattices in the chiral limit, however the data 
suggest that an even finer lattice is desirable. The large momentum
behavior is examined to determine the quark condensate.
\end{abstract}


\maketitle

\section{Introduction}
The quark propagator is one of the fundamental quantities in QCD. By studying 
the momentum-dependent quark mass function in the infrared region we can gain 
valuable insight into the mechanism of dynamical chiral symmetry breaking and 
the associated dynamical generation of mass.  
There have been several studies of the momentum space quark 
propagator~\cite{soni1,soni2,jon1,jon2,Bow02a,Bow02b,blum01,overlgp,overlgp2} 
in Landau gauge using different fermion actions.  Here we focus on the 
overlap fermion action and extend previous work~\cite{overlgp} to three 
lattices with different lattice spacing $a$ at fixed physical volume.  This 
allows us to study the approach of the Landau gauge quark propagator to the 
continuum limit.  The study of the overlap quark propagator in the Gribov copy 
free Laplacian gauge is underway and will be reported elsewhere.

\section{Quark Propagator on the Lattice}
\label{lattice}

In a covariant gauge in the continuum, the renormalized Euclidean
space quark propagator has the form
\begin{eqnarray}
S(\zeta^2;p)=\frac{1}{i {p \slh} A(\zeta^2;p^2)+B(\zeta^2;p^2)}
=\frac{Z(\zeta^2;p^2)}{i{p\slh}+M(p^2)}\, ,
\label{ren_prop}
\end{eqnarray}
where $\zeta$ is the renormalization point. The renormalization 
point boundary conditions are chosen to be
\begin{equation}
Z(\zeta^2;\zeta^2)\equiv 1 \, , \qquad M(\zeta^2)\equiv m(\zeta^2) \, .
\end{equation}
where $m(\zeta^2)$ is the renormalized quark mass an ultraviolet 
renormalization point.
The functions $A(\zeta^2;p^2)$ and $B(\zeta^2;p^2)$, or alternatively
$Z(\zeta^2;p^2)$ and $M(p^2)$,  contain all of the nonperturbative information
of the quark propagator.  Note that $M(p^2)$ is renormalization
point independent, {\it i.e.}, since $S(\zeta^2;p)$ is multiplicatively
renormalizable all of the renormalization-point dependence is carried by
$Z(\zeta^2;p^2)$.  For sufficiently large momenta the effects of dynamical
chiral symmetry breaking become negligible, {\it i.e.}, for large $p^2$ and we have
$M(p^2)\to m(\zeta)$ up to logarithmic corrections, where $m(\zeta)$ is 
the perturbative running mass.

When all interactions for the quarks are turned off, {\it i.e.}, when the gluon
field vanishes (or the links are set to one), the quark propagator has its 
tree-level form
\begin{equation}
S^{(0)}(p)=\frac{1}{i{p\slh}+m^0} \, ,
\end{equation}
where $m^0$ is the bare quark mass.  When the interactions with the
gluon field are turned on we have
\begin{equation}
S^{(0)}(p) \to S^{\rm bare}(a;p) = Z_2(\zeta^2;a) S(\zeta^2;p) \, ,
\label{tree_bare_ren}
\end{equation}
where $a$ is the regularization parameter - in this case, the lattice spacing 
- and $Z_2(\zeta^2;a)$ is the quark wave-function renormalization constant
chosen so as to ensure $Z(\zeta^2;p^2)=1$.  For simplicity of notation we 
suppress the $a$-dependence of the bare quantities.

On the lattice we expect the bare quark propagators, in momentum space,
to have a similar form as in the continuum, except
that the $O(4)$ invariance is replaced by a 4-dimensional hypercubic
symmetry on an isotropic lattice.
Hence, the inverse lattice bare quark propagator takes the general form
\begin{equation}
(S^{\rm bare})^{-1}(p)\equiv
{i\left(\sum_{\mu}C_{\mu}(p)\gamma_{\mu}\right)+B(p)}.
\label{invquargen}
\end{equation}
We use periodic boundary conditions in the spatial directions
and anti-periodic in the time direction.
The discrete momentum values for a
lattice of size $N^{3}_{i}\times{N_{t}}$, with $n_i=1,..,N_i$ and $n_t=1,..,N_t$, are
\begin{eqnarray}
p_i=\frac{2\pi}{N_{i}a}\left(n_i-\frac{N_i}{2}\right),\hspace{1cm}{\rm{and}}\hspace{1cm}p_t
=\frac{2\pi}{N_{t}a}\left(N_t-\frac{1}{2}-\frac{N_t}{2}\right).
\label{dismomt}
\end{eqnarray}

The overlap fermion formalism~\cite{neuberger0,neuberger2}
realizes an exact chiral
symmetry on the lattice and is automatically ${\cal O}(a)$ improved.
The massive overlap operator can be written as~\cite{edwards2}
\begin{eqnarray}
D(\mu) = \frac{1}{2}\left[1+\mu+(1-\mu)\gamma_5 \epsilon(H_w) \right] \, ,
\label{D_mu_eqn}
\end{eqnarray}
where $H_w(x,y)=\gamma_5 D_w(x,y)$ is the Hermitian Wilson-Dirac
operator, the mean-field improved Wilson-Dirac operator can be written as
\begin{eqnarray}
D_{w}(x,y)
&=& \left[ (-m_wa)+4r\right]\delta_{x,y}
-\frac{1}{2}\sum_\mu\left\{(r-\gamma_\mu)
U_\mu(x)\delta_{y,x+\hat\mu}+(r+\gamma_\mu)
U^\dagger_\mu(x-a\hat{\mu})\delta_{y,x-\hat\mu}\right\} \nonumber \\
&=& \frac{u_0}{2\kappa} \left[\delta_{x,y}
-\kappa\sum_\mu
\left\{(r-\gamma_\mu)
\frac{U_\mu(x)}{u_0}\delta_{y,x+\hat\mu}+(r+\gamma_\mu)
\frac{U^\dagger_\mu(x-a\hat{\mu})}{u_0}\delta_{y,x-\hat\mu}
\right\}
\right] \, .
\label{diracop}
\end{eqnarray}

The negative Wilson mass $(-m_wa)$ is then related to $\kappa$ by
\begin{equation}
\kappa\equiv\frac{u_0}{2(-m_wa)+(1/\kappa_c)} \, ,
\label{kappa_defn}
\end{equation}
and mean-field improvement allows the use of the tree-level value
$\kappa_c = 1/(8 r)$.  The Wilson parameter is typically chosen to be
$r=1$ and we will also use $r=1$ here in our numerical simulations.
The dimensionless quark mass parameter is
\begin{equation}
\mu \equiv \frac{m^0}{2m_w} \, .
\label{mu_defn}
\end{equation}
The overlap quark propagator is given by the equation
\begin{equation}
S^{\rm bare}(m^0)\equiv \tilde{D}_c^{-1}(\mu) \, ,
\label{overlap_propagator}
\end{equation}
where 
\begin{equation}
\tilde{D}_c^{-1}(\mu) \equiv \frac{1}{2m_w} \tilde{D}^{-1}(\mu)
\hspace{1cm}{\rm{and}}\hspace{1cm}
\tilde{D}^{-1}(\mu) \equiv \frac{1}{1-\mu}\left[{D}^{-1}(\mu)-1\right]
\, .
\label{D_mu}
\end{equation}

At tree-level, the inverse bare lattice quark propagator becomes the 
tree-level version of \eref{invquargen}
\begin{equation}
(S^{(0)})^{-1}(p)\equiv
{i\left(\sum_{\mu}C_{\mu}^{(0)}(p)\gamma_{\mu}\right)+B^{(0)}(p)}\,.
\label{treeinvpro}
\end{equation}
We calculate $S^{(0)}(p)$ directly by setting the links to unity in the
coordinate space quark propagator and taking its Fourier transform.
It is then possible to identify the appropriate kinematic lattice
momentum directly from the definition
\begin{equation}
q_\mu\equiv C_{\mu}^{(0)}(p)
\label{latmomt}
\end{equation}
The form of $q_\mu(p_\mu)$ is shown and its analytic form given in 
Ref.~\cite{overlgp}.  Having identified the appropriate kinematical lattice 
momentum $q$, we can now define the bare lattice propagator as
\begin{equation}
S^{\rm bare}(p)
\equiv \frac{Z(p)}{i{q\slh}+M(p)}.
\end{equation}
This ensures that the free lattice propagator is identical to the free 
continuum propagator.  Due to asymptotic freedom the lattice propagator will
also approach the continuum form at large momentum.  In the gauge sector, this
analysis approach dramatically improves the gluon 
propagator~\cite{lei99, gluon_refs}.

The two Lorentz invariants can now be \footnote{This is merely an
illustrative example.  See Ref.~\cite{overlgp} for details on how these 
functions are actually calculated.}
\begin{gather}
Z^{-1}(p) = \frac{1}{12iq^2} \tr \{q\slh S^{-1}(p) \} \\
M(p) = \frac{Z(p)}{12} \tr \{ S^{-1}(p) \}.
\end{gather}
While $Z$ is directly dependent on our choice of momentum, $q$, the
mass function $M$ is indirectly dependent on this choice.  In the case
of staggered quarks it has been seen that the kinematic momentum
derived from tree-level analysis of the action is a good choice of
momentum for the mass function~\cite{Bow02a,Bow02b}.  This is an
empirical result.  The tree-level behavior of the Overlap quark
propagator is rather different, however, and a different approach may
be needed.  We investigate this issue by analyzing the scaling
behavior of the propagator over three values of the lattice spacing at
constant physical volume.

\section{NUMERICAL RESULTS}
\label{numerical}

We present results from three lattice ensembles, each with a different lattice 
spacing, $a$, but having the same physical volume. Lattice parameters are summarized in 
Table~\ref{simultab}. The gauge configurations were
created using a tadpole improved plaquette plus rectangle 
(L\"{u}scher-Weisz~\cite{Lus85}) gauge action.  Each ensemble consists of 
50 configurations.  The lattice spacing was determined by the static quark 
potential using the string tension $\sqrt{\sigma}=440$~MeV~\cite{zanotti}.

\begin{table}[ht]
\caption{\label{simultab}Lattice parameters.}
\begin{ruledtabular}
\begin{tabular}{cccccccc}
Action &Volume &$N_{\rm{Therm}}$ & $N_{\rm{Samp}}$ &$\beta$ &$a$ (fm) & $u_{0}$
 & Physical Volume (fm$^4$)\\
\hline
Improved       & $16^3\times{32}$ & 5000 & 500 & 4.80 & 0.093  & 0.89650 & 
$1.5^3\times{3.00}$ \\
Improved       & $12^3\times{24}$ & 5000 & 500 & 4.60 & 0.123  & 0.88888 & 
$1.5^3\times{3.00}$ \\
Improved       & $8^3\times{16}$  & 5000 & 500 & 4.286& 0.190  & 0.87209 & 
$1.5^3\times{3.00}$ \\
\end{tabular}
\end{ruledtabular}
\end{table}

The gauge field configurations were gauge-fixed to the \oa{2} improved Landau 
gauge~\cite{bowman2}.  Our calculation begins with the evaluation of the
inverse of the Dirac operator in Eq.~(\ref{D_mu_eqn}).  We approximate the 
matrix sign function $\epsilon(H_w)$ by a 14th-order Zolotarev 
approximation~\cite{zolo}.  The coordinate space propagator, 
Eq.~(\ref{overlap_propagator}), is calculated for each configuration.  
A discrete Fourier 
transform is then applied to the each of the coordinate space propagator, and 
the momentum-space bare quark propagator, $S^{\rm bare}(p)$, is finally obtained from 
the ensemble average.

In the Wilson action we use $\kappa=0.19163$ for the 
regulator mass.  We calculate the Overlap quark propagator for ten quark
masses on each ensemble by using a shifted Conjugate Gradient solver.  The 
quark mass parameter $\mu$ was adjusted to make the tree level bare quark mass in physical
units, the same on three lattices.  For example, we choose 
$\mu = 0.018,$ 0.021, 0.024, 0.030, 0.036, 0.045, 0.060, 0.075, 0.090, and 
0.105 on ensemble 1, {\it i.e.}, the $16^3\times{32}$ lattice with $a$ = 0.093 fm. 
This corresponds to bare masses in physical units of
$m^0 = 2 \mu m_w = $ $127$,
$148$, $169$, $211$, $254$, $317$,
$423$, $529$, $634$, and $740$~MeV respectively.


Results from ensemble 2 were presented in Ref.~\cite{overlgp}, and some
results from ensemble 3 were also reported in Ref.~\cite{overlgp2}. 
Here we will compare the quark propagators on each ensemble to examine its
behavior as the lattice spacing vanishes.
First we present some results from ensemble 1, the finest lattice of the three.
All data has been cylinder cut~\cite{lei99}.  Statistical uncertainties are
estimated via a second-order, single-elimination jackknife.

\begin{figure}[tp]
\centering\includegraphics[width=8.0cm,angle=90]{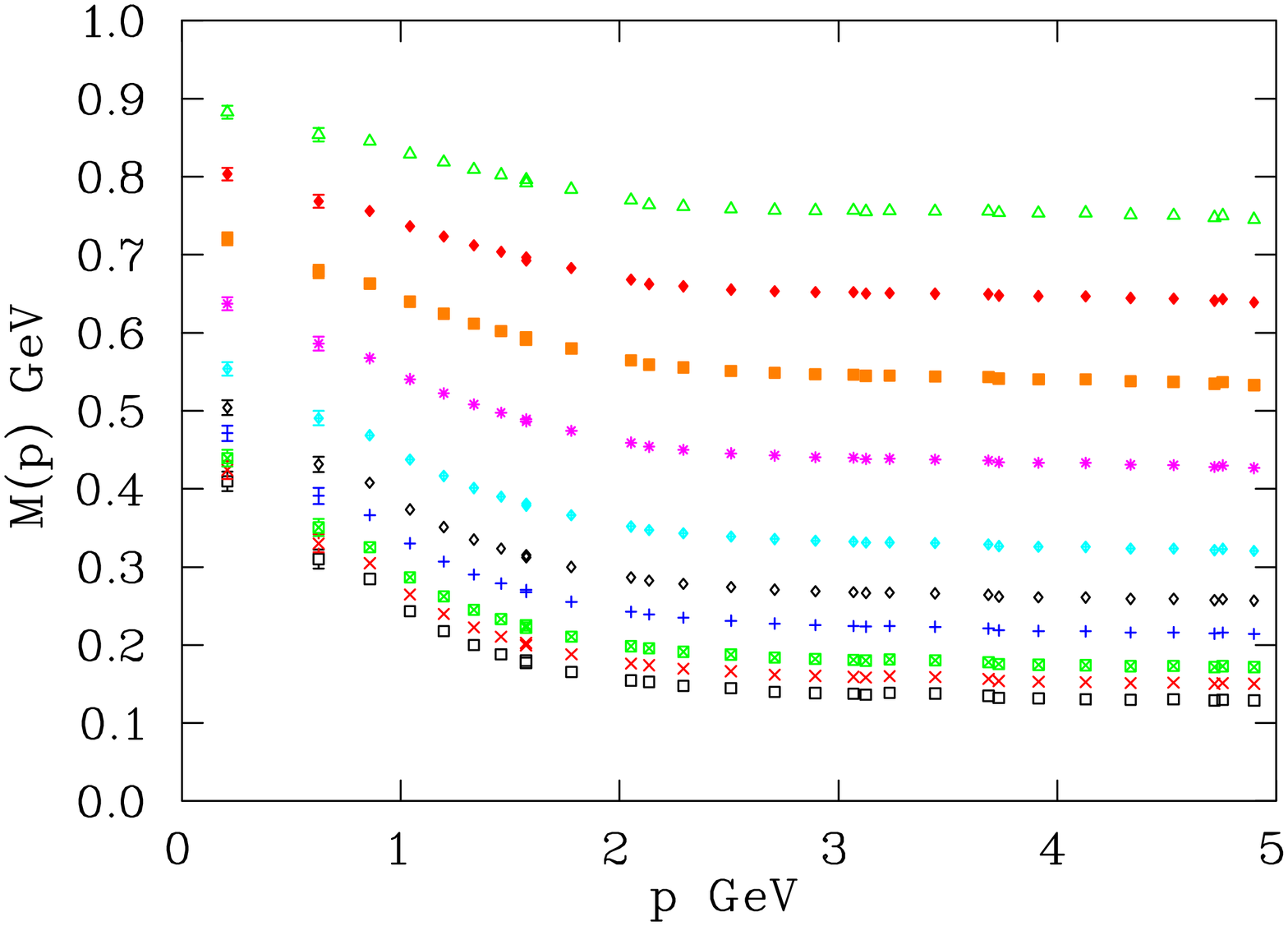}
\centering\includegraphics[width=8.0cm,angle=90]{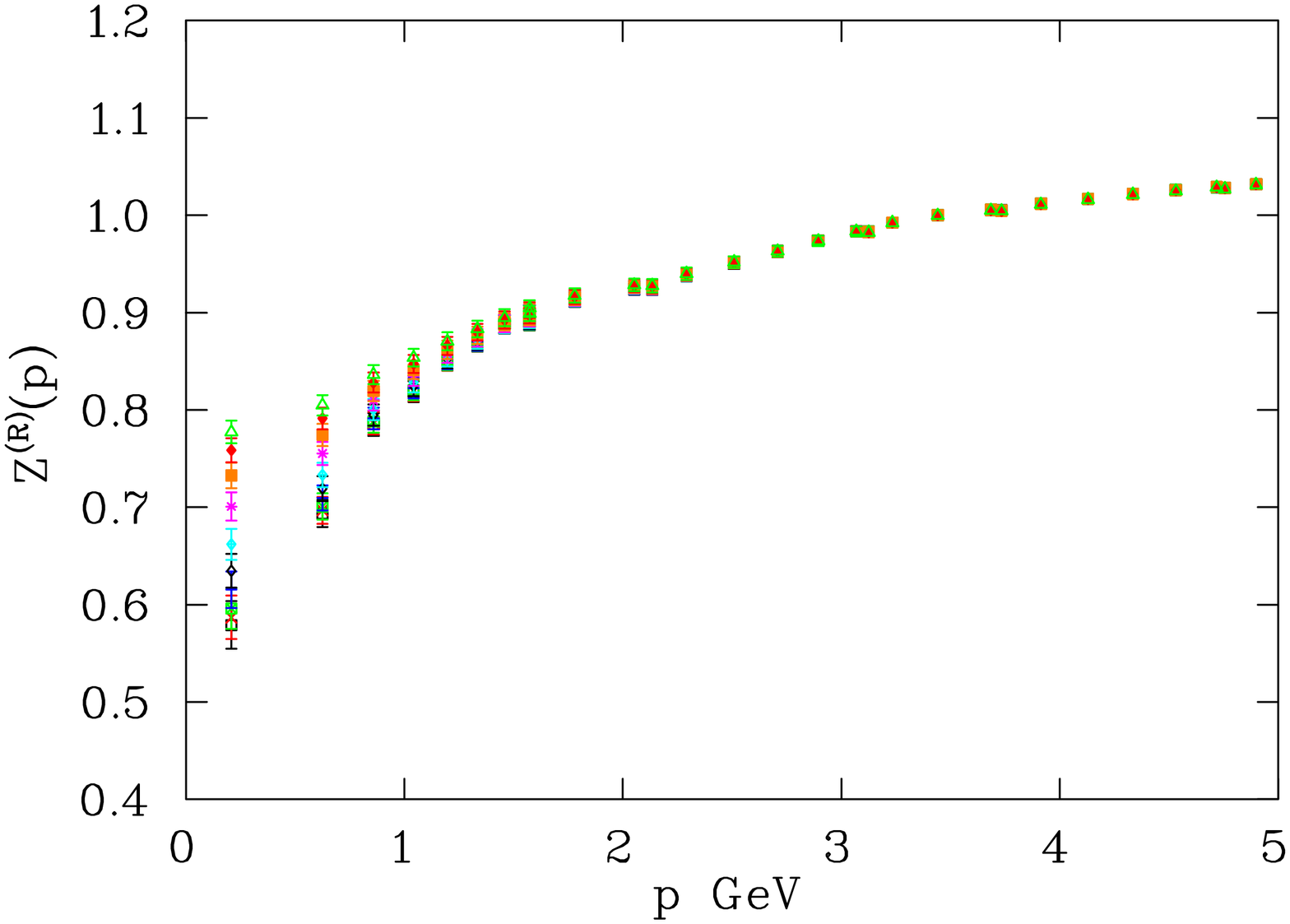}
\caption{The functions $M(p)$ and $Z^{(\rm{R})}(p)\equiv Z(\zeta^2;p)$  
renormalized at $\zeta=3.44$~GeV for all ten quark masses.  Data are plotted 
versus the discrete momentum values defined in Eq.~(\ref{dismomt}),
$p=\sqrt{\sum{p_\mu^{2}}}$, over the interval [0,5] GeV.
The data correspond to bare quark masses (from bottom to top) 
$\mu = 0.018,$ 0.021, 0.024, 0.030, 0.036, 0.045, 0.060, 0.075, 0.090, and 
0.105, which in physical units corresponds to $m^0 = 2\mu m_w \simeq $ $127$, 
$148$, $169$, $211$, $254$, $317$,
$423$, $529$, $634$, and $740$~MeV respectively.
\label{combmovrp}}
\end{figure}

In Fig.~\ref{combmovrp} we show the results for all ten masses for both the 
mass and wave function renormalization functions,
$M(p)$ and $Z^{(\rm{R})}(p)\equiv Z(\zeta^2;p)$ respectively, as a function of
the discrete lattice momentum $p$.  $Z^{(\rm R)}(p)$ is renormalized at 
$\zeta=3.44$~GeV.  We see that both $M(p)$ and $Z^{(\rm R)}(p)$ are reasonably
well-behaved up to 5~GeV.
In the plots of $M(p)$ the data is ordered as one would expect by the values 
for $\mu$, {\it i.e.}, the larger the bare quark mass $m^0$ the higher is the $M(p)$ 
curve.  At small bare masses $M(p)$ falls off more rapidly with momentum, 
which is understood from the fact that a larger proportion of the infrared mass
is due to dynamical chiral symmetry breaking at small quark masses.  
In the non-relativistic limit the mass function would be a constant.
 
\begin{figure}[tp]
\centering\includegraphics[width=8.0cm,angle=90]{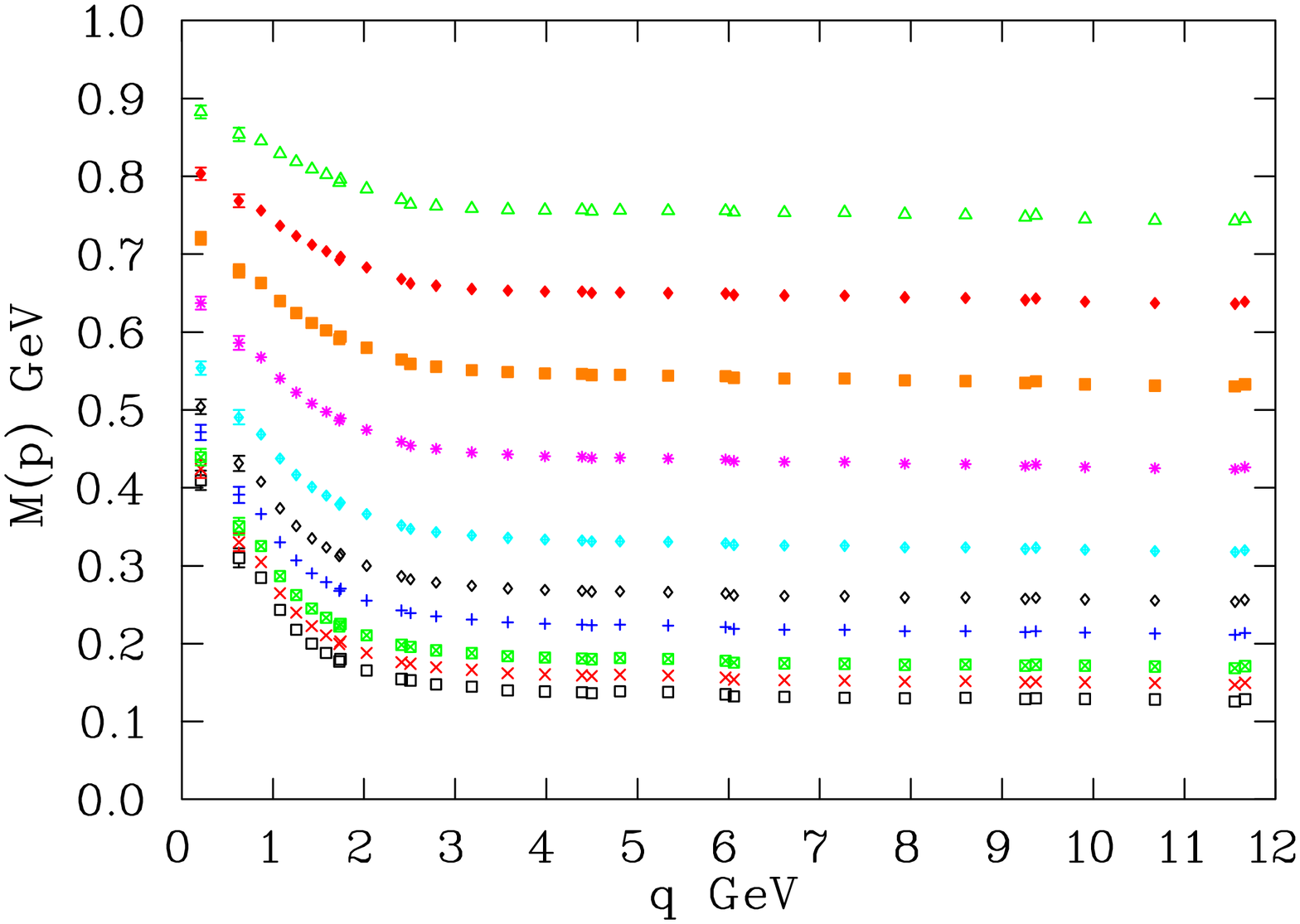}
\centering\includegraphics[width=8.0cm,angle=90]{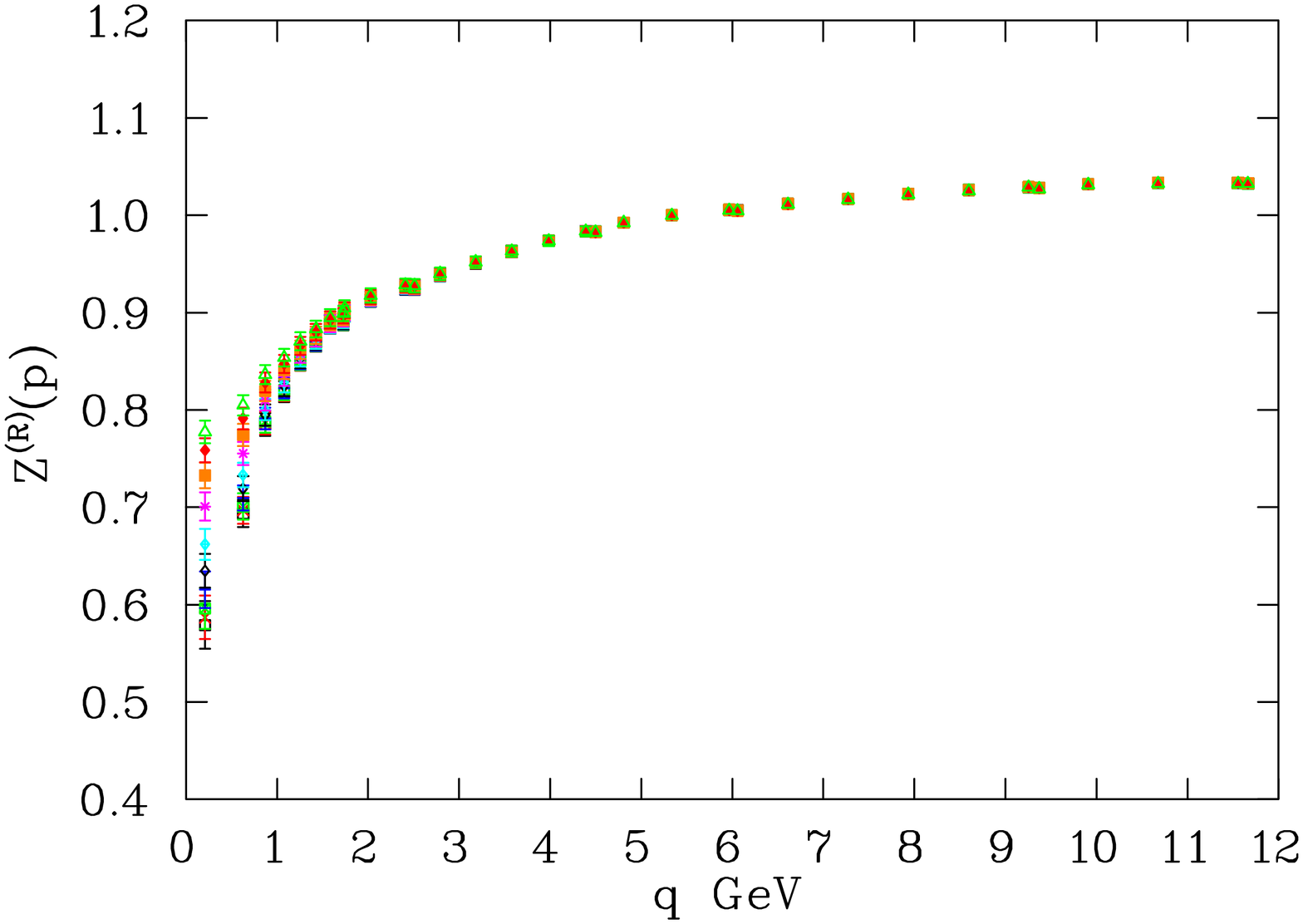}
\caption{The functions $M(p)$ and $Z^{(\rm{R})}(p)\equiv Z(\zeta^2;p)$ for 
renormalization point $\zeta $ =5.31~GeV for all ten quark masses.  
Data are shown versus the discrete momentum values defined in 
Eq.~(\ref{latmomt}), $q=\sqrt{\sum{q_\mu^{2}}}$, over the interval [0,12] GeV.
The data in both parts of the figure correspond from
bottom to top to increasing quark masses.  The values of the bare
quark masses are in the caption of Fig.~\protect{\ref{combmovrp}}.
\label{combmovrq}}
\end{figure}

$Z^{(\rm R)}(p)$ on the other hand is infrared suppressed.  The smaller the 
quark mass, the more pronounced the dip at low momenta.  This behavior is 
qualitatively consistent with what is seen in Dyson-Schwinger based QCD 
models\cite{agw94,Alkofer}.  It is likely that some of the suppression, however,
is due to the finite volume~\cite{Bow02b}.
In Fig.~\ref{combmovrq} we repeat these plots but now using the kinematical 
lattice momentum $q$.  This only alters the large momentum behavior of the
propagator.

%
%
\begin{figure}[tp]
\centering\includegraphics[width=8.0cm,angle=90]{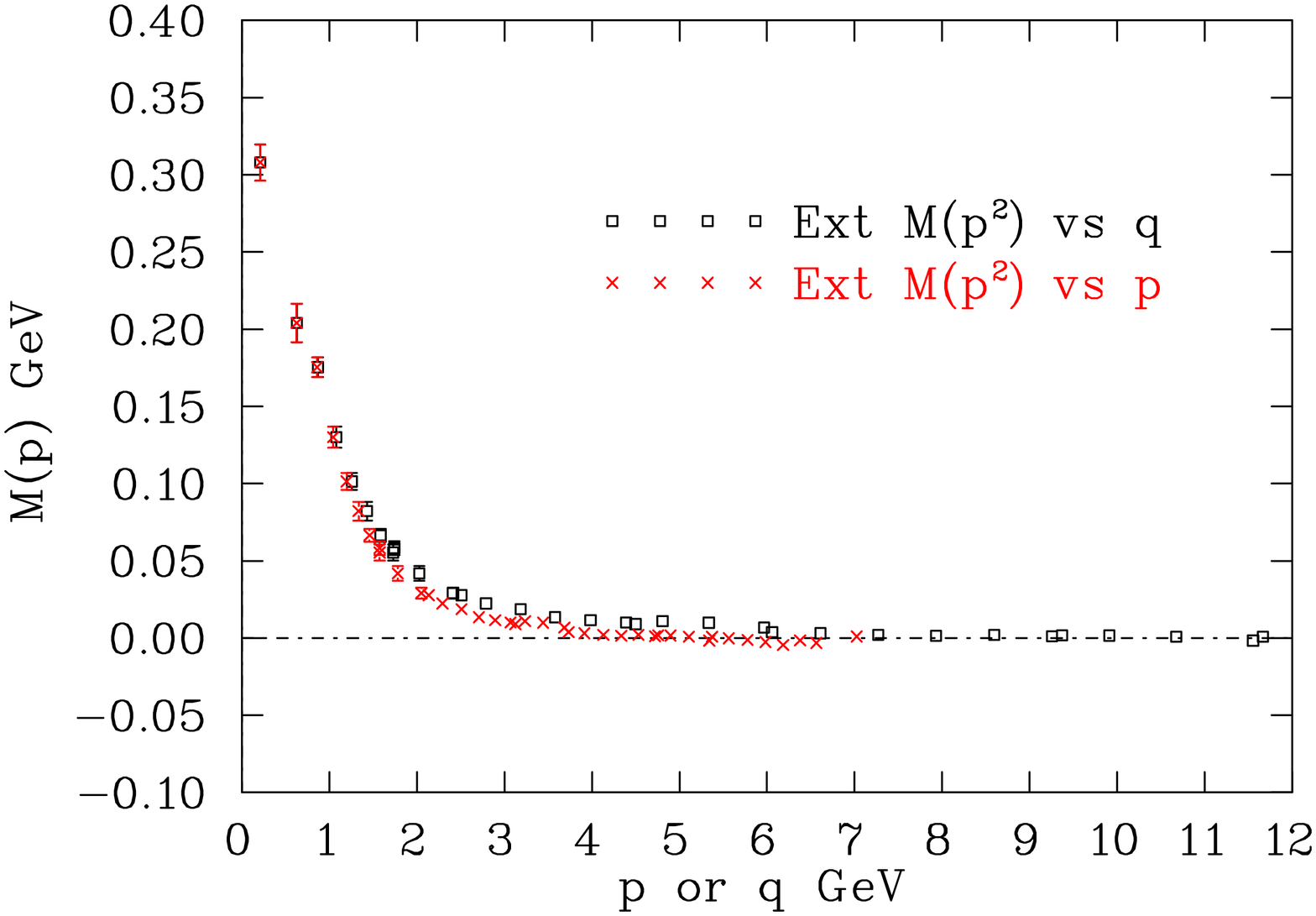}
\centering\includegraphics[width=8.0cm,angle=90]{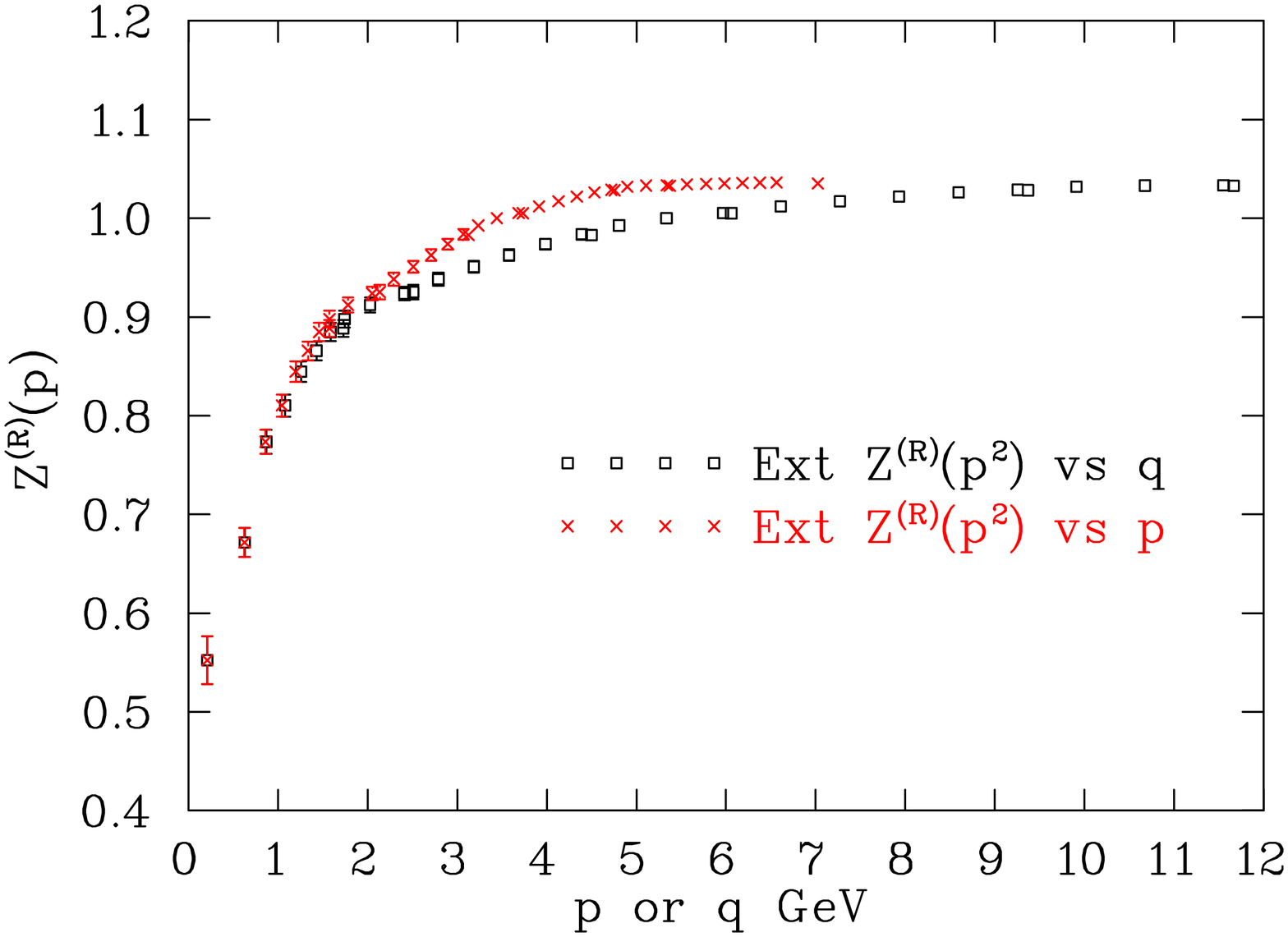}
\caption{Linearly extrapolated estimates of $M(p)$ and $
Z^{(\rm R)}(p)\equiv Z(\zeta^2;p)$ in the chiral limit. Here the renormalization point
are $\zeta $ =3.44~GeV  in the $p$ scale and $\zeta $ =5.31~GeV in the $q$ scale.
At the smallest accessible momentum $M_{\text{IR}} = 307(6)$~MeV and 
$Z_{\text{IR}} = 0.55(2)$.
\label{extramz2pi}}
\end{figure}

We perform an extrapolation to zero quark mass by a linear fit to the data. 
At sufficiently large momenta the mass function will be proportional
to the bare quark mass, in which case the linear extrapolation is appropriate.
Non-linear behavior is to be expected in the infrared, but this simple ansatz
describes the present data adequately.  In the ultraviolet, the renormalized 
$Z$ should - and
does - approach its perturbative value of 1.  This is mass independent.
We investigated the effect of including the smallest quark masses in 
the chiral extrapolation and found that eliminating the lightest two made 
little difference to the extrapolated result. 
The resulting estimate of the chiral limit is shown in Fig.~\ref{extramz2pi}.
These are shown against both $p$ and $q$, renormalized as before.
We see that both $M(p)$ and $Z^{(\rm R)}(p)$ deviate strongly from their
tree-level behavior.  In particular, as in earlier studies of the Landau 
gauge quark propagator\cite{jon1,jon2,Bow02a,overlgp}, we find a clear signal 
of dynamical mass generation and a significant infrared suppression of the 
$Z^{(\rm R)}(p)$ function.  At the most infrared point - the lowest non-zero 
momentum available on this lattice - the dynamically generated mass has the 
value $M_{\rm IR}=307(6)$~MeV and the momentum-dependent wave function
renormalization function has the value $Z_{\rm IR}=0.55(2)$.
These values are very similar to the results found in previous
studies~\cite{jon1,jon2,Bow02a,Bow02b,overlgp} and are also similar to
typical values in QCD-inspired Dyson-Schwinger equation
models~\cite{agw94,Alkofer}.  The results of Ref.~\cite{Bow02b} suggests that
at least some of the infrared suppression of $Z^{(\rm R)}(p)$ is due to finite 
volume effects.

\begin{figure}[tp]
\centering{\epsfig{angle=90,figure=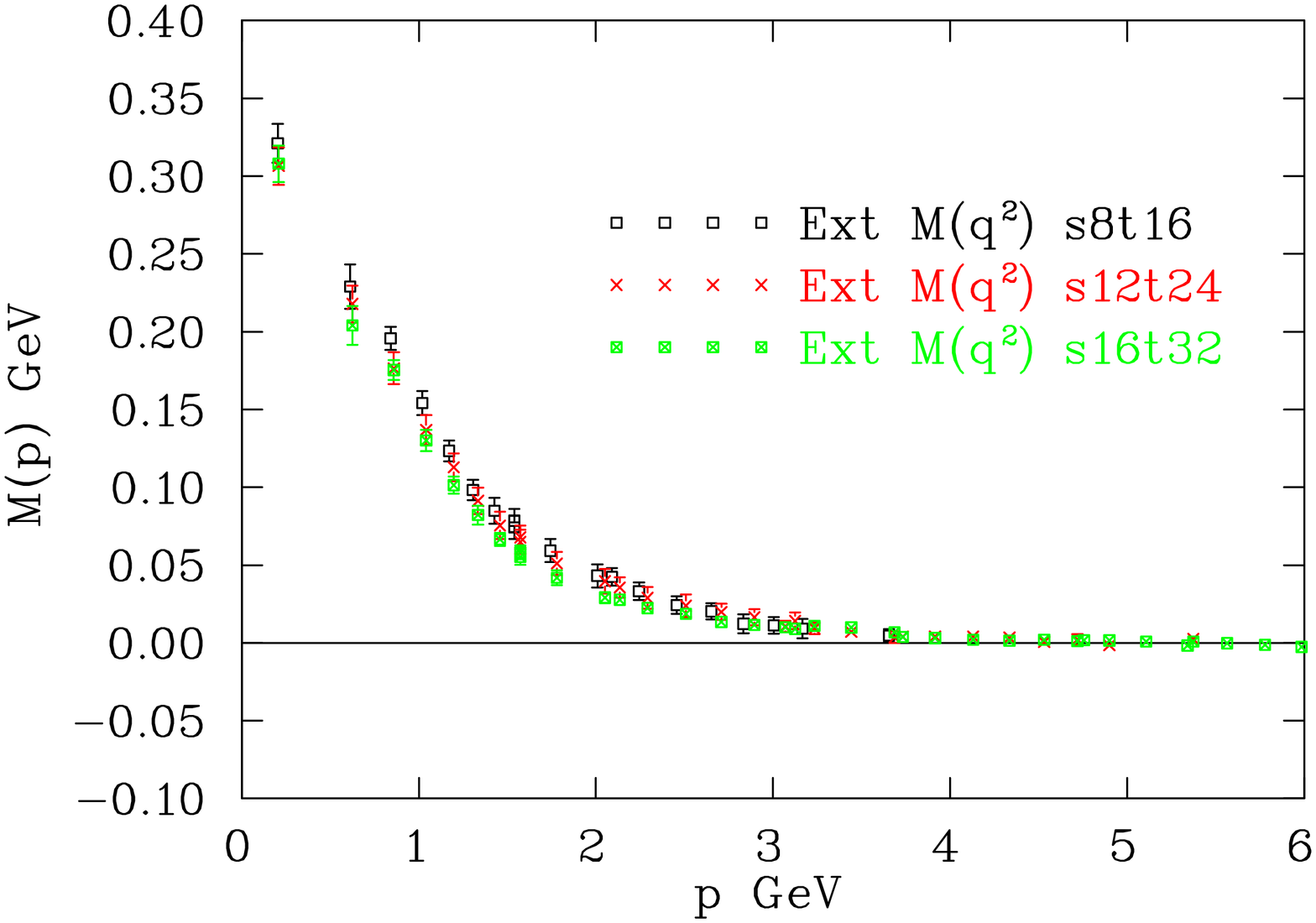,height=8cm} }
\centering{\epsfig{angle=90,figure=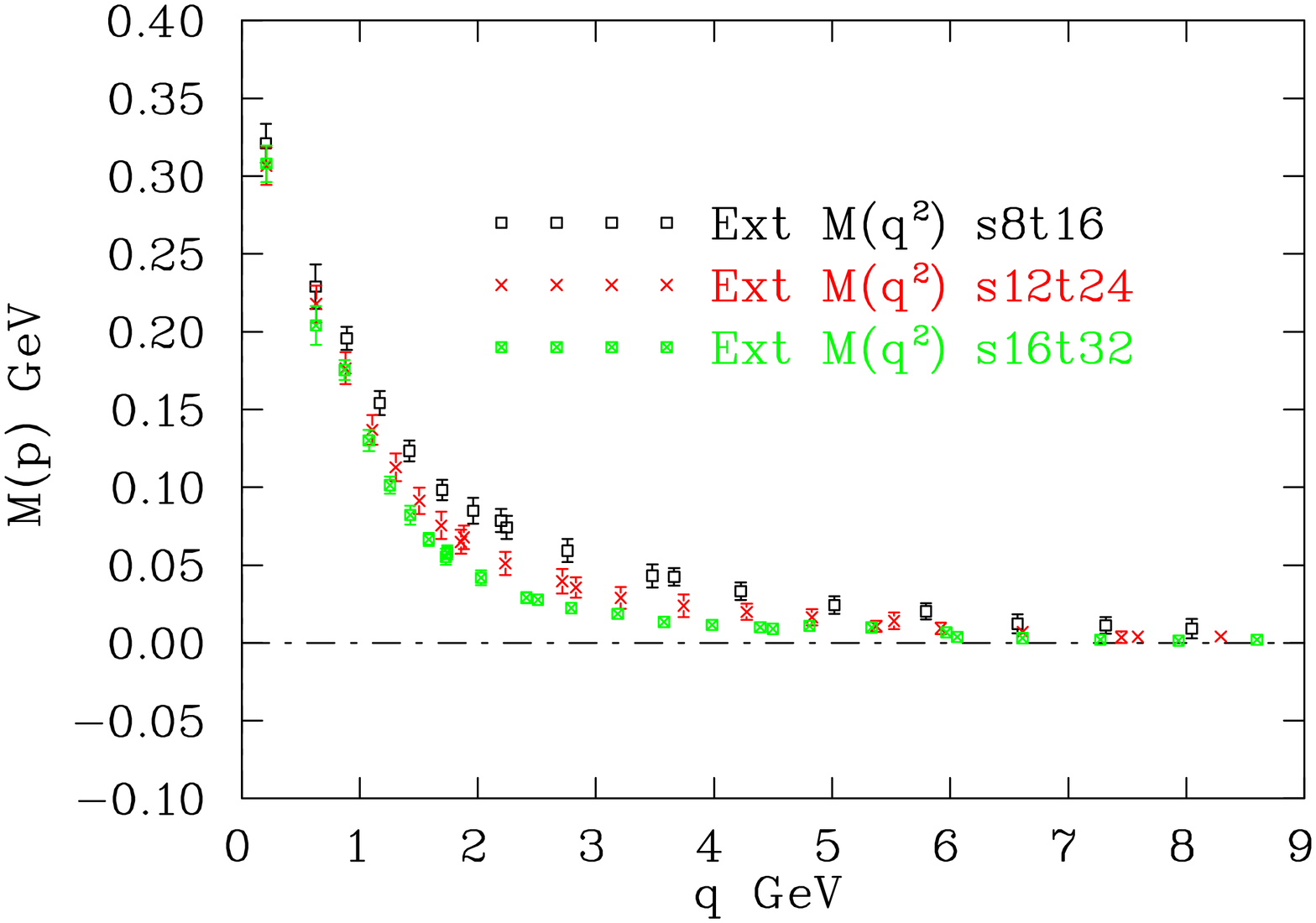,height=8cm} }
\caption{The mass function $M(p)$ from a linear extrapolation to the chiral
limit is shown for our three lattices.  
In the upper part of the figure $M(p)$ is plotted against the discrete lattice
momentum $p$, whereas in the lower part it is plotted against the kinematical
momentum $q$.  The results again suggest that we most rapidly approach the 
continuum limit by plotting $M(p)$ against $p$.}
\label{compare_cm}
\end{figure}

   Now we present the results on three lattices for comparison.  These
lattices have approximately the same physical volume, but each has a 
different lattice spacing.  Thus we can study the Overlap propagator's scaling 
properties.  We present the results for the chiral limit.  The mass function, 
$M(p)$ for the three lattices in the chiral limit is plotted in 
Fig.~\ref{compare_cm}, using both $p$ and $q$.  We see that if the mass 
function $M(p)$ is plotted against the standard lattice momentum $p$, the 
agreement of the results among the three lattices is better than the case in 
which the mass function $M(p)$ is plotted against kinematical lattice momentum 
$q$.

\begin{figure}[tp]
\centering{\epsfig{angle=90,figure=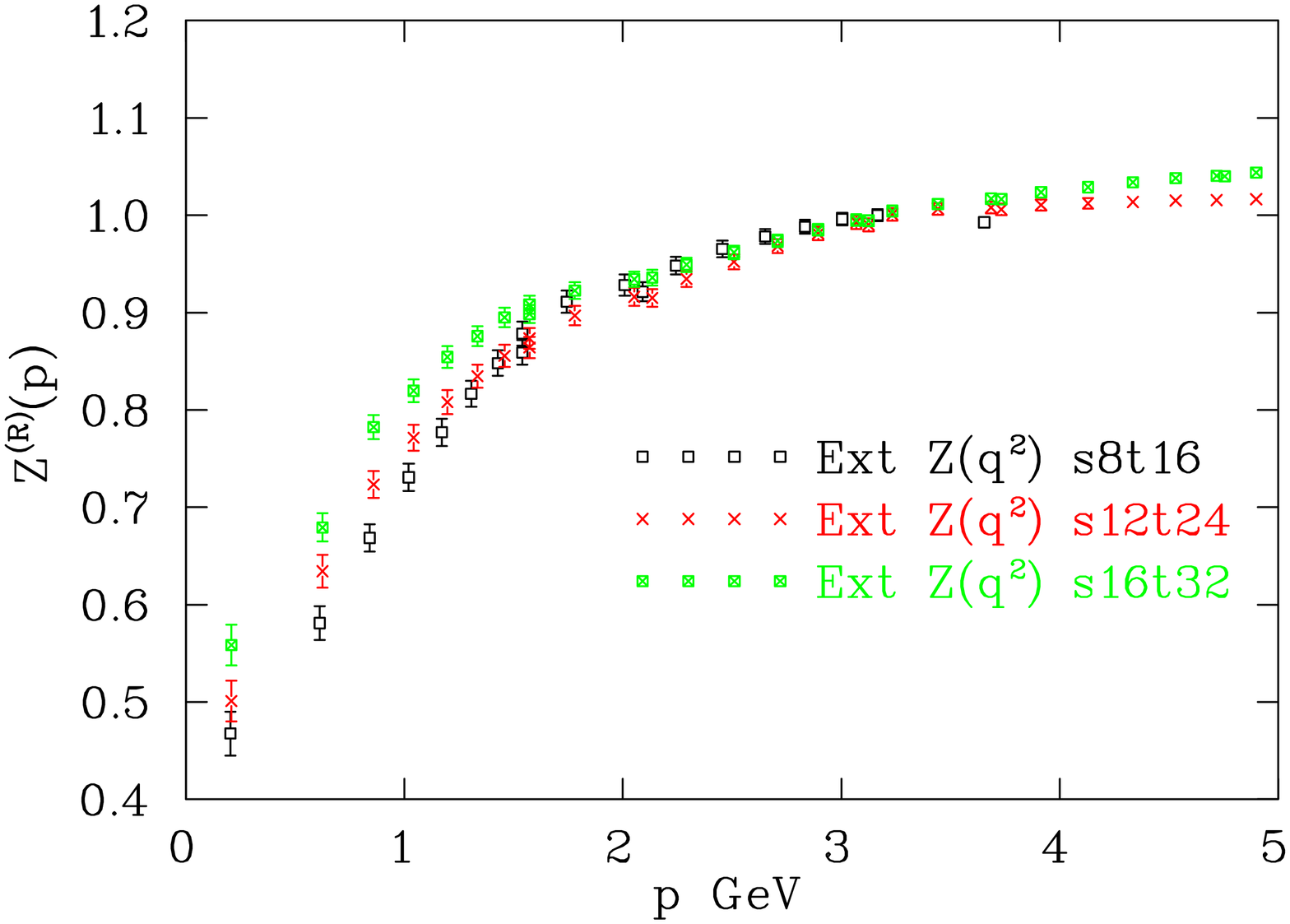,height=8cm} }
\centering{\epsfig{angle=90,figure=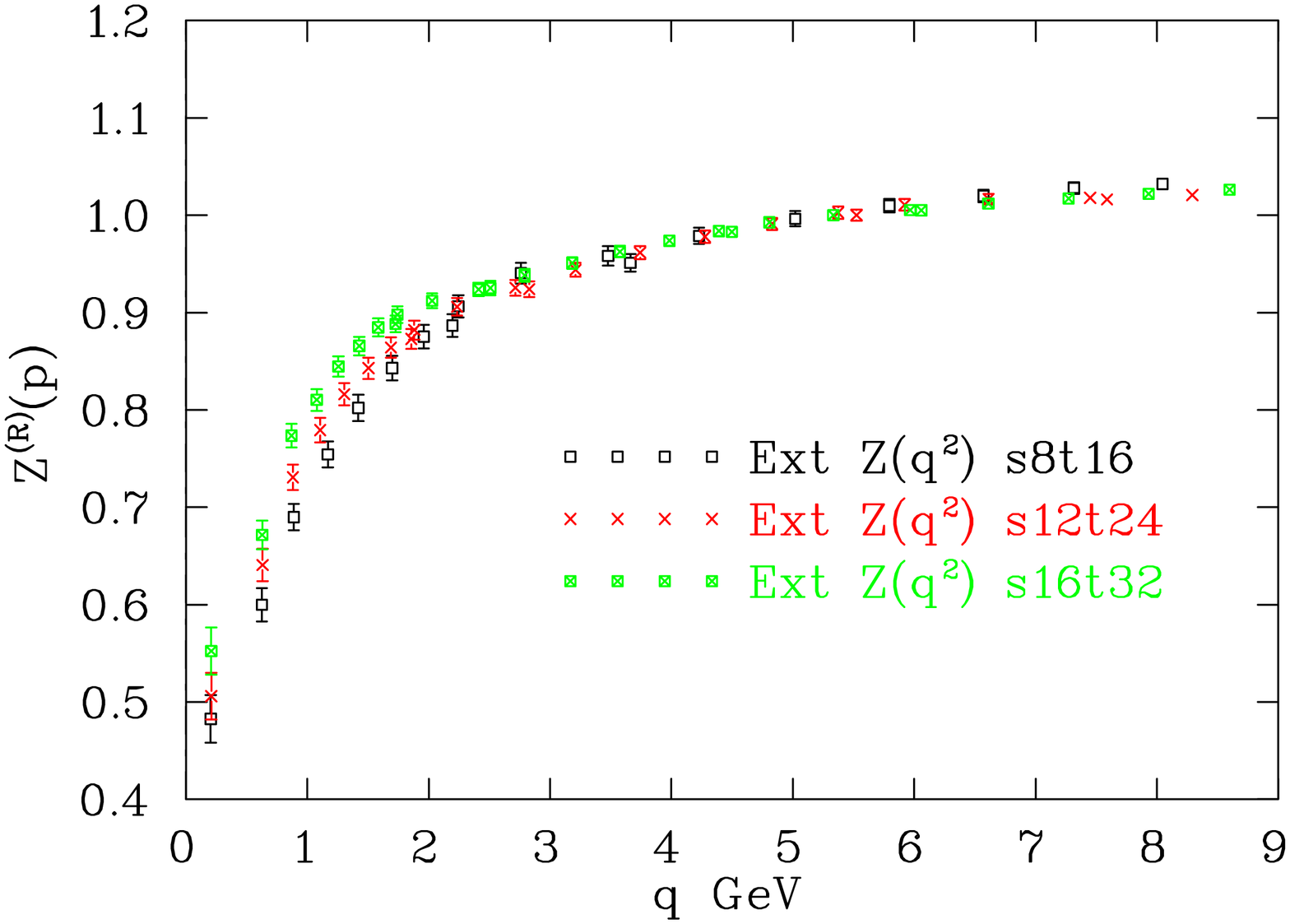,height=8cm} }
\caption{The momentum-dependent wave function renormalization function 
$Z^{(\rm{R})}(p)\equiv Z(\zeta^2;p)$ for renormalization point $\zeta=3.44$~GeV
in the $p$-scale and 
$\zeta=5.31$~GeV in the $q$-scale from a linear extrapolation to the chiral 
limit.  In the upper part of the figure
$Z^{(\rm{R})}(p)$ is plotted against the discrete lattice momentum $p$ whereas
in the lower part it is plotted against the kinematical momentum $q$.  The
results again suggest that we most rapidly approach the continuum limit by 
plotting $Z^{(\rm{R})}(p)$ against $q$.}
\label{compare_cz}
\end{figure}

The results for the renormalization function 
$Z^{(\rm R)}(p) \equiv Z(\zeta^2;p)$ of the three lattices is plotted in 
Fig.~\ref{compare_cz}.  Here the renormalization point are chosen to be
$\zeta =$ 3.44 GeV in $p$ scale and $\zeta =$ 5.31 GeV in $q$ scale.  Contrary 
to the case of mass function $M(p)$,  but as predicted by the tree-level 
analysis, the agreement between the results on three lattices is better if 
$Z^{(\rm R)}(p)$ is plotted against the kinematical lattice momentum $q$.  
There are also the relatively small discrepancies in $Z^{(\rm R)}(p)$ versus
$q$ in the infrared region on three lattices, it seems that in the continuum 
limit, the dip in the renormalization function $Z^{(\rm R)}(p)$ will be narrow 
but the depth of the dip will be unchanged.  It suggests that an even finer 
lattice will be needed to confirm the continuum limit of $Z^{(\rm R)}(p)$ in 
the infrared.  It is possible that the linear chiral extrapolation is 
unreliable for $Z^{(\rm R)}(p)$  in this regime or it could be that dynamical 
chiral symmetry breaking is coupling hypercubic lattice artifacts
to finite volume effects. This warrants further investigation with finer and 
larger lattices.

Thus we have resolved one of the key questions raised in the studies of 
Ref.~\cite{overlgp}. We see that the continuum limit appears to be 
approached most rapidly when $Z^{(\rm R)}(p)$ is plotted against $q$ and 
$M(p)$ is plotted against $p$.  The better scaling of $Z^{(\rm R)}(p)$ as a 
function of $q$ is natural and predicted by the tree-level analysis.  That 
$M(p)$ is better against $p$ is purely observation.

Another way of studying scaling is by making comparisons with known continuum
results.
In the chiral limit, in the continuum, the asymptotic quark mass function has
the form
\begin{equation}
\label{eq:mass_asymp}
M(p^2) \stackrel{p^2\rightarrow \infty}{=} - \frac{4\pi^2 d_M}{3}
   \frac{\langle\psibar \psi\rangle}{[\ln(\mu^2 / \Lqcd^2)]^{d_M}}
   \frac{[ \ln(p^2 / \Lqcd^2) ]^{d_M-1}}{p^2}
\end{equation}
(see Ref.~\cite{agw94}, Eq.~(6.15)) where the anomalous
dimension of the quark mass is $d_M = 12/(33 - 2N_f)$ for $N_f$ quark
flavors ($N_f = 0$ in the present case).  The dependence of $M(p^2)$ on the
renormalization point $\mu$ is canceled by the dependence of the
condensate, maintaining the
renormalization point invariance of the mass function.  We fit this form
to the lattice data obtained by both linear and quadratic chiral extrapolation.
A quadratic extrapolation was used on the AsqTad data~\cite{Bow02b} so it
is useful for comparing with those results.  Some results are presented in
Table~\ref{tab:condensate}.

\begin{table}[h]
\caption{\label{tab:condensate} Extracted values of the quark condensate $\langle\psibar \psi\rangle $.}
\begin{ruledtabular}
\begin{tabular}{ll|ll|ll}
 $\beta$ &extrapolation & $p$ fit region (GeV) & $\cond^{1/3}$ (MeV) & $q$ fit region (GeV) & $\cond^{1/3}$ (MeV)  \\
\hline
   4.60  &   linear     &    3.6 - 4.5     & 337(39) & 4.3 - 8.6  & 621(49) \\
   4.60  &   quadratic  &    3.6 - 4.5     & 292(56) & 4.3 - 8.6  & 575(72) \\
   4.80  &   linear     &    3.6 - 5.3     & 327(22) & 5.5 - 11.4 & 499(34) \\
   4.80  &   quadratic  &    3.6 - 5.3     & 259(36) & 5.5 - 11.4 & 395(54) 
\end{tabular}
\end{ruledtabular}
\end{table}

The difference between the quadratic and linear extrapolations is no great
surprise as our quark masses are rather heavy.  This is a constraint of the
volume.  The relevant point is that there is good agreement between the two
lattices when $p$ is the momentum, but not when $q$ is the momentum.

\section{summary and outlook}

The momentum space quark propagator has been calculated in Landau gauge on 
three lattices with different lattice spacing $a$ but very similar physical
volumes in order to explore the approach to the continuum limit.
We calculated the nonperturbative momentum-dependent wave function 
renormalization $Z(\zeta^2;p)$ and the nonperturbative mass function $M(p)$ for a 
variety of bare quark masses.  We also explored the quark propagator in  
the chiral limit.
As previously anticipated~\cite{overlgp}, the continuum
limit for $Z(\zeta^2;p)$ is approached most rapidly when it is plotted against the
kinematical lattice momentum $q$, whereas for the quark mass function, $M(p)$,
we have found that using the discrete lattice momentum $p$ provides the most 
rapid approach to the continuum limit. 

Future work should test our conclusions and further explore the continuum
limit with one or more additional finer lattice spacings.  In addition, it
will be necessary to use both finer and larger volume lattices, in particular to study 
the infrared behavior of $Z(\zeta^2,q)$.
One can also use other kernels in the overlap fermion formalism,
e.g., using a fat-link irrelevant clover (FLIC) action~\cite{Zanotti:2001yb}
as the overlap kernel~\cite{Kamleh,Kusterer} in order to further establish the
robustness of our conclusions and to provide more accurate data.
These studies are currently underway and results will be reported elsewhere.

\section{Acknowledgments}

Support for this research from the Australian Research Council is gratefully 
acknowledged. Supercomputing support from the Australian
National Facility for Lattice Gauge Theory, the South Australian
Partnership for Advanced Computing (SAPAC) and the Australian
Partnership for Advanced Computing (APAC) is gratefully acknowledged.

\end{document}